\documentclass{sig-alternate}

\usepackage{booktabs} % For formal tables
\usepackage{multirow}
\usepackage{subfigure}

\usepackage[ruled]{algorithm2e} % For algorithms

\SetAlFnt{\small}
\SetAlCapFnt{\small}
\SetAlCapNameFnt{\small}
\SetAlCapHSkip{0pt}
\IncMargin{-\parindent}

\usepackage{url}
\usepackage{hyperref}

\usepackage{color}
\definecolor{Orange}{rgb}{0.9,0.5,0}
\definecolor{NavyBlue}{rgb}{0.1, 0.4, 0.8}
\definecolor{Magenta}{rgb}{0.8, 0.1, 0.6}
\definecolor{Green}{rgb}{0.1, 0.8, 0.3}
\definecolor{DarkGreen}{rgb}{0.0, 0.7, 0.2}
\definecolor{Brown}{rgb}{0.4, 0.3, 0.1}
\definecolor{Burgundy}{rgb}{0.5, 0.0, 0.13}
\definecolor{BrightCerulean}{rgb}{0.11, 0.67, 0.84}

\newcommand{\numnotifs}[1]{78,930~}

\newcommand{\compresslist}{
  \setlength{\itemsep}{3pt}
  \setlength{\parskip}{0pt}
  \setlength{\parsep}{0pt}
}

\begin{document}
% Title portion
\title{Continual Prediction of Notification Attendance with Classical and Deep Network Approaches}

\numberofauthors{2}
\author{
\alignauthor
Kleomenis Katevas\thanks{Part of the work was done during a student internship at Telef\'onica Research. All authors contributed equally.}\\
       \affaddr{Cognitive Science Research Group}\\
       \affaddr{Queen Mary University of London, UK}\\
       \email{k.katevas@qmul.ac.uk}
% 2nd. author
\alignauthor
Ilias Leontiadis, Martin Pielot, Joan~Serr{\`a}\\
       \affaddr{Telef{\'o}nica Research, Barcelona, Spain}\\
       \email{name.surname@telefonica.com}
}

\maketitle

\begin{abstract}
We investigate to what extent mobile use patterns can predict -- at the moment it is posted -- whether a notification will be clicked within the next 10 minutes.
We use a data set containing the detailed mobile phone usage logs of 279~users, who over the course of 5~weeks received 446,268~notifications from a variety of apps.
Besides using classical gradient-boosted trees, we demonstrate how to make continual predictions using a recurrent neural network (RNN).
The two approaches achieve a similar AUC of ca. 0.7 on unseen users, with a possible operation point of 50\% sensitivity and 80\% specificity considering all notification types (an increase of 40\% with respect to a probabilistic baseline).
These results enable automatic, intelligent handling of mobile phone notifications without the need for user feedback or personalization.
Furthermore, they showcase how forego feature-extraction by using RNNs for continual predictions directly on mobile usage logs.
To the best of our knowledge, this is the first work that leverages mobile sensor data for continual, context-aware predictions of interruptibility using deep neural networks.

\end{abstract}

% We no longer use \terms command
\terms{Human Factors}


\section{Introduction} \label{sec:intro}
% ================================================================== 

Notifications are alerts that try to attract attention to new content \cite{Iqbal2010}, such as unread emails or social network activity. 
While they help us to stay on top of our messages, emails, or social network activity, they annoy and disrupt \cite{Pielot2014notif}, impair work performance even when they are not attended~\cite{Stothart2015}, and negatively affect well-being~\cite{Kushlev:2016} as they induce symptoms of hyperactivity and inattention. At the same time, notifications are essential for people to keep up with social expectations of responding in a timely manner~\cite{Pielot:2017}. Therefore, the research community is investigating ways to reduce the negative effects of notifications by predicting how open or reactive a user would be to a notification prior to its delivery, in order to enable intelligent ways of handling them~\cite{Turner:2015}.

Previous research on related problems has employed classical, feature-based machine learning, in which features have to manually be derived from sensor events. Choosing features is a creative, manual, and inherently time-consuming task. 
Consequently, during feature extraction, important information may not be modeled and thus remain unused by the classifier~\cite{Choy2016}. 

Deep learning proposes to solve this problem \cite{Goodfellow16BOOK}:
a cascade of neural network layers is employed, where each subsequent layer can learn more complex information. The model implicitly identifies and learns predictive `features' from the data and tasks at hand.
Deep networks are significantly outperforming state-of-the-art methods in several domains, such as image-~\cite{Krizhevsky12NIPS}, text-~\cite{manning2016computational}, or speech-~\cite{Graves2013SpeechRW} processing. These domains deal with complex input data that presents important spatial regularities. However, data from mobile phone usage logs is irregular and yields sparse vectors. For example, hours or just seconds may pass between two unlocks of the phone. This irregularity poses a significant challenge for neural networks. Recurrent neural networks (RNNs)~\cite{Goodfellow16BOOK} effectively deal with variable-length sequential data but, nonetheless, they are typically designed for constant rate, synchronous sequences~\cite{Lee:2016:PhasedLSTM}. 
Thus, it is an open question whether such networks can perform at par with classical, feature-based machine learning approaches when feeding raw sensor data.

% IN THIS PAPER
We report from an investigation into how to model users' attentiveness to notifications from mobile phone sensor data. On the basis of data collected from 279~mobile phone users over the course of 5~weeks, we created machine-learning models to perform continual predictions for each user.
We both employ a classical machine learning approach and develop a new deep learning pipeline to leverage raw mobile phone sensor data. 
% CONTRIBUTIONS
The main contributions of the article are:
\begin{itemize}\compresslist

    \item evidence that real time mobile phone data can predict, prior to posting a notification, whether the user will launch the associated application within 10 minutes. We do so with a mean AUC of 0.7, 40\% above the baseline.
    
    \item a practical pipeline for processing mobile sensor data and usage events for continual deep learning predictions. To address the fact that mobile sensor events can be sparse, we propose a scheme to compress the data, resulting into 5.1\% prediction improvement, 20x data-size reduction, and 28x faster learning speeds. This pipeline may be extensible to other mobile sensing applications using deep neural networks.
    
    \item evidence that, in the given case of notification attendance, an RNN using raw data performs at par with (if not better than) a competitive, feature-based classifier.
    
\end{itemize}

% \clearpage

\section{Related Work} \label{sec:related}
% ==================================================================

\subsection{Notification Attendance}
\label{related_work_notif_attendance}

% Iqbal and Bailey \cite{Iqbal2008} define a notification as a visual cue, auditory signal, or haptic alert generated by an application or service that relays information to a user outside her current focus of attention.
Mobile phone users receive dozens or even hundreds of notifications per day~\cite{Pielot2014notif, Sahami2014}, and typically attend to them within minutes~\cite{Pielot2014notif}. 
\emph{Attending} a notification refers to any observable reaction to a notification, ranging from simply triaging and dismissing it, to opening the corresponding application.
% The majority of those notifications originates from communication channels, primarily messengers and email \cite{Sahami2014} .
Mobile phone notifications frequently interrupt other activities~\cite{Pielot2014notif}. 
When a person has been interrupted, it takes time to switch back to the original task~\cite{Czerwinski2004}. Notifications are even disruptive when they are not attended~\cite{Stothart2015}, and a large number of notifications may further have negative effects on mental well-being~\cite{Kushlev2016}.
% Pielot \emph{et al.} \cite{Pielot2014notif} discuss different options of using such information to improve the user experience: (1) suppressing alerts in the case of unimportant notifications (as proposed in \cite{Rosenthal2011}), (2) delaying notifications until an opportune moment arises (as proposed in \cite{Horvitz2005b}), or (3) communicating unavailability in the case of computer-mediated communication (as proposed in \cite{Pielot2014}). 
Therefore, an increasing number of studies explore ways to reduce or infer the \emph{disruptiveness} of notifications or the level of \emph{interruptibility} of mobile phone users. Such studies can be classified into three groups: 
%\begin{enumerate} \compresslist

    %\item 
    \textbf{Self-reported interruptibility} --- The first approach is to estimate interruptibility from self-reports collected via experience sampling and other artificial interruptions~\cite{Choy2016,Dali2015,Fischer2011,Pejovic2014,Rosenthal2011,Turner2017}. For example, Pejovic and Musolesi~\cite{Pejovic2014} proposed a platform for triggering experience-sampling notifications at opportune moments. A study app triggered 8~notifications per day, which asked participants to report whether the current moment was a good moment for an interruption.

    %\item 
    \textbf{Monitoring reaction to study-generated notifications} --- 
    The second approach is measuring the reaction to notifications that are generated as part of a study \cite{Fischer2010,Pejovic2014,Poppinga2014,Sarker2014,Okoshi2015,Okoshi2017,Pielot2015,Pielot2017,Rosenthal2011}. 
    For example, Poppinga \emph{et al.} \cite{Poppinga2014} explored predicting opportune moments to remind users to self-report their mood. On the basis of 6,581~notifications from 79~active users, they predicted with an AUC of 0.72 whether a notification would be opened. 
    Okoshi \emph{et al.}~\cite{Okoshi2017} tested a breakpoint-detection system to time notifications of Yahoo! Japan. Based on data from over 680,000 users, they show that mean response times can be significantly reduced from 54 to 27~minutes. Increases in click-through rates and engagement scores were found too, but not statistically significant. 
    %In our previous work
     %In a real-world deployment of the boredom classifier developed as part of that work, they showed that people who are predicted to be bored are more open to notification-delivered suggested articles. Compared to the control group, click-through rates were 2.5~times higher.
    
    %\item 
    \textbf{Monitoring reaction to existing phone notifications} --- 
    The third and most closely-related approach is measuring the reaction to notifications created by actual applications~\cite{Mehrotra2015,Mehrotra2016,Pielot2014}. %The present article falls under this category, hence we discuss the related work in detail.
    Mehrotra \emph{et al.}~\cite{Mehrotra2015} explored interruptibility with respect to notification content, its sender, and the context in which a notification is received. For 3~weeks, 35~volunteers installed a study app, which collected sensor data as well as subjective data from 6-item questionnaires each day. They found that messages from family members and emails from work-related contacts have the highest acceptance rate (defined by attending to the notification within 10 minutes). % Other types of notifications are largely `rejected' as per the definition above. 
    In a follow-up study, Mehrotra \emph{et al.}~\cite{Mehrotra2016} analyzed 474 self-reports related to previously received notifications and found that people primarily click on notifications in a timely fashion because the sender is important, or because its content is important, urgent, or useful. % When a notification is disrupting, its content becomes more decisive than the sender. 
    Pielot \emph{et al.}~\cite{Pielot2014} showed that contextual data purely derived from the mobile phone, such as screen activity or hour of the day, can predict with a precision of 81.2\% whether a notification from a messaging application, such as WhatsApp, will be attended within 6.15~minutes (median time until a notification was attended). % However, they did not explore other types of notifications.

%\end{enumerate}

\vspace*{0.3cm}\noindent
\textbf{What is missing} --- According to a survey by Turner \emph{et al.}~\cite{Turner:2015}, ``little progress has been made on generalised approaches for interruptibility''. 
Results from study-generated notifications may not generalize to other use cases.
Self-reports may be biased, \emph{e.g.}, as people may not realize in what situations they actually react to notifications.
The few studies where predictions on actual notifications are made are either confined to the quite specific category of messaging, or require data which cannot yet be derived in an automated way.
Thus, our analysis of how 279 users reacted to 446,268 notifications fills the gap of showing to what extent a generalized intelligent system can, without requiring user input, predict whether a user will engage with an application shortly after it posted a notification.
% With an analysis of 446,268 notifications from 279 users, our study further stands out in terms of magnitude of the analyzed data set.

\subsection{Deep Networks for Mobile Sensor Data}
% ----------------------------------------------------

%The use of deep neural networks is a modern machine learning approach that is rapidly gaining popularity across the research community. One of the main advantages is that it allows to set up classifiers capable of learning the feature set automatically from almost raw labeled, or even unlabeled data, in a supervised way, without the need of developing a series of hand-crafted features per study. A cascade of neural network layers is employed, where each subsequent layer can learn more complex information, typically in a hierarchical fashion. The model implicitly identifies and learns predictive 'features’ from the available dataset and for the task at hand.

%Deep neural networks have been outperforming traditional machine learning classification tasks, especially in the areas of computer vision [add cites], speech recognition [add cites] and natural language processing [add cites]. 
 %Human activity recognition in one example that deep learning outperformed other state-of-the-art methods \cite{Ronao16ESA}. \minos{lots of papers that we can add here} 

In the last few years, the use of deep neural networks (DNNs) has reached the mobile sensing community. %, allowing the automated creation of the feature set using raw mobile sensor data as an input. 
%\noindent
Ronao and Cho~\cite{Ronao16ESA} used deep convolutional networks with time-series sensor data (accelerometer and gyroscope) for detecting the human activity with 94.79\% accuracy. Lane \emph{et al.}~\cite{Lane15UBICOMP} presented a solution focused on audio sensing tasks (ambient scene analysis, stress detection, emotion recognition, and speaker identification) that outperformed other state-of-the-art classifiers designed for mobile devices in both accuracy and robustness to acoustic diversity.
%
% Old version of the previous paragraph:
% Lane et al.~\cite{Lane15UBICOMP} presented DeepEar, a mobile audio sensing framework that uses DNNs to perform common audio sensing tasks. The framework outperformed popular datasets (that is, EmotionSense, StressSense, SpeakerSense, JigSaw) in both accuracy and robustness to acoustic diversity compared to other state-of-the-art classifiers specialized in that domain. 
%

%\noindent
DeepSense~\cite{Yao:2016:DeepSense} is, to our knowledge, the first published work that inputs a wide range of time-series mobile sensor data into a DNN. In such work, the attendance to large time spans is achieved by using a combination of convolutional and RNN layers. The framework outperformed the baselines in a number of tasks, namely car tracking, activity recognition, and user identification. Even though the authors did not report any results with a wide range of mobile sensors, they claim that their framework can be directly applied to almost all other sensors, such as microphone, Wi-Fi signal, barometer, and light sensor.
%

%\noindent
On the basis of data from 18,000 Emotion Sense users, Servia \emph{et al.}~\cite{Servia2017} present evidence regarding the extent to which valence can be inferred from mobile phone sensor data. By using a restricted Boltzmann machine, they were able to distinguish between positive and negative valence with an accuracy of 62\% (week days) and 68\% (weekends). 
%Unfortunately, the paper does not disclose the size of the positive and negative classes, so it is impossible to put this performance in context.
%

%\noindent
Felbo \emph{et al.}~\cite{Felbo16ICLRW} proposed a hybrid approach of both traditional machine learning and DNNs for predicting demographics from mobile phone meta-data (number of unique contact, calls, text messages, etc.),
%. They trained a convolutional network that produces the features, and later input them to a support vector machine for predicting the age and gender of the participants. Interestingly,
showing the hybrid approach outperformed the DNN in both age and gender prediction.

\vspace*{0.3cm}\noindent
\textbf{What is missing} --- These previous works largely rely on data from non-sparse data sources. Past work in interruptibility prediction, however, showed that sparse events, such as screen activity or app launches, are often strong predictors. This work addresses the open research question of how to address the sparsity-issue when feeding these events into RNNs.

% In the context of predicting openness to interruptions, previous work so far has only employed classical feature-based models. The open research question is to deal with the sparsity of some events, such as app launches or screen activity, which were found to be highly predived in past work.

% With this study, we explore to what extent deep neural networks can be employed in this use case instead of classical, feature-based, machine learning approaches. Since almost no prior art exists, we also develop a pipeline detailing all necessary steps for using mobile sensor data in deep neural networks.

\section{Methodology}  \label{sec:method}
% ==================================================================

The present work has two primary goals. The first goal is to investigate to what extent mobile phone usage data allows to predict, prior to posting it, whether a notification will be consumed timely. Therefore, we analyze a data set that contains detailed phone usage logs of 279~Android phone users.
We create a feature-based classifier to predict whether the application associated with a notification will be opened within the next 10~minutes. 
Such a prediction enables a wide range of applications, such as suppressing notifications that are not likely to be clicked, delaying notifications until opportune moments or, in the context of communication, communicate to the sender whether a timely response is likely or not.

The second goal is to explore whether and how DNNs can be applied to raw mobile usage logs, and whether they can perform as good as the classical, feature-based approaches. We therefore create an RNN-based classifier to perform the same predictions as the feature-based one, and we present both a quantitative and a qualitative comparison between the two.

It is worth mentioning that, while some mobile phone usage logs used in this study correspond to actual\linebreak hardware-based sensors (\emph{e.g.}, Accelerometer, Light, etc.), some others correspond to software-based, or virtual sensors (\emph{e.g.}, App, Notification, Screen, etc.). For simplicity, in the rest of this work, we will refer to all of them as \emph{mobile phone sensors}.

\subsection{Data Set}
% -------------------------------------------------------------

The data set was collected during a dedicated study conducted in summer 2016. Participants were recruited through a specialized agency. We requested a sample that matches the gender and age distribution of the country of study in Western Europe. The only restriction was that people were required to own an Android phone. Android phones account for the large majority (approximately 90\%) of the smartphone share in the country of study. For this analysis, we consider participants who owned a phone with Android OS Version 18 or higher. The reason behind this decision is that this version introduced the \emph{NotificationListener} service. This allows users to grant applications access to all notification events which was required by this study.

People with interest in joining the study were first directed to the informed consent, which had been approved by the legal department of our institution. The consent form listed all data to be collected and gave extra details about potential personally-identifiable information. The participants were then taken to an installation guide that explained how to install the app. During this process, participants registered the app to listen for notification and accessibility events. We ran informal usability tests to ensure that the installation process was fast and easy to understand. 

The data collection commenced once the app was installed, set-up properly, and once the participants confirmed their agreement with the informed consent from within the app. 
The app kept running in the background for the duration of the study while passively collecting rich sensor data about the user's context and phone usage (see below). The data set contains 279~participants (147~female, 132~male). The self-reported ages range from 18 to 66 ($M = 37.70$, $SD = 11.05$). On average, the duration of the participation was 26.17 days ($SD = 7.21$). The data set contains 446,268~notifications. Table~\ref{tab:num_notifs} details, by category, a number of data set statistics.

\begin{table}[t]
    \setlength{\tabcolsep}{5pt}
    \centering
    \begin{tabular}{l|r|rr|r}
    \hline\hline
    \bf Category & \bf Count & \multicolumn{2}{c|}{\bf User \& day} & \bf Positive \\ 
    \bf          &      & \bf Mean & \bf SD & \bf cases (\%) \\ 
    \hline
    Messaging          & 244726 & 35.7 	& 37.3  & 59.1 \\
    Email              & 76075 	& 17.7 	& 23.6  & 17.5 \\
    Productivity       & 57972 	& 9.4 	& 14.3  & 21.1 \\
    Social             & 41008 	& 10.0 	& 21.5  & 25.5 \\
    Entertainment      & 11824 	& 5.1 	& 9.0   & 21.2 \\
    Games              & 10435 	& 7.2 	& 11.8  & 24.8 \\
    Alert              & 4228 	& 3.0 	& 5.5   & 12.5 \\
    \hline\hline
    \end{tabular}
    \caption{Data set statistics. From left to right: total number of notifications per category, statistics per user and day, and fraction of positive cases (notifications that triggered an app launch in less than 10~minutes).}
    \label{tab:num_notifs}
\end{table}

\subsection{Classical Approach with Gradient Boosted Trees} 
\label{sec:xgboost}
% ==================================================================

\subsubsection{Pipeline}
\label{pipeline}
% -------------------------------------------------------------
% Feature extraction

Mobile sensor data can be categorized into periodical, where sensors regularly report data (\emph{e.g.}, accelerometer, light, etc.), and event-driven, where new data are reported when an event occurs (\emph{e.g.}, the screen is unlocked, a notification is received, etc.). Some of the periodical data (listed in Table~\ref{tab:sensors}) can be seen as features (\emph{i.e.} Mean and Max Linear Acceleration, Mean of Noise, Semantic Location) and not as raw sensor data. This data was not a result of the feature extraction process. Instead, the app already provided this information in internally-aggregated form for power efficiency, and in some cases for respecting the participant's privacy. Moreover, this sensor data is basic and easy to be computed, in contrast with the more advanced features we explain below.

The feature-extraction process is inspired by the work of Pielot \emph{et al.}~\cite{Pielot2017}, where they predict whether people are open to engagement via notification-delivered content. Our goal was to characterize the moment (and indirectly the context) before a notification was posted. We computed features corresponding to three different time windows: the \emph{current moment} (last 5~minutes), \emph{recent} (last hour), and \emph{current day} (since 5\,am today). For each of these time windows, and in line with related work, we compute 145 features that belong to 5~different groups of variables: \emph{Communication Activity}, \emph{Context}, \emph{Demographic}, \emph{Phone Status} and \emph{Usage Patterns}. Our general strategy was to introduce as many features as possible from all the available sensors and let the classifier automatically identify the most indicative ones. Table~\ref{tab:sensors} describes the sensors of our study. We selected these sensors on the basis of the results from \cite{Pielot2017}, choosing those that provided the most predictive power.

For each of the sensors, we kept the last-reported value as a feature (current moment). For continuous sensors that report numerical values, we extracted the quartiles and the median absolute deviation during the three time windows reported above. For the case of the semantic location sensor, we created features that contain the category of the location (home, work, out, unknown; one-hot encoded), the percentage of time spent in those categories (one-hot encoded), and the number of different locations visited during the day. Regarding the app usage, we created features to model the time since the last app was launched, the number of apps launched during the three time windows, and the category of the last opened application (alert, email, entertainment, games, messaging, productivity, social, system). For notifications, we created features to model the time since the last notification was received; the number of notifications received and the category of the last received notification (alert, email, entertainment, games, messaging, productivity, social, other). For the rest of the event-driven sensors, we aggregated the number of times the sensor has reported data (\emph{e.g.}, number of device unlocks, screen orientation changes) during the three time windows. Independently from sensors, we created features corresponding to the age and the gender of the user. In addition, we included the hour of the day, day of the week, as well as whether the current day is a working day or not (including public holidays) of the moment when the notification was posted.

% Minos: -4ex is required to fix centering due to p{13cm}
\begin{table*}[t]
\small
\centering
\begin{tabular}{|c|l|p{12.2cm}|}
    \hline
    & Sensor             & Description \\
    \hline \hline
    \multirow{6}{*}[-1.5ex]{\rotatebox[origin=c]{90}{Periodical}}
    & Accelerometer      & Mean and maximum linear acceleration. \\
    & Battery            & Percentage of the device's battery drain per hour. \\
    & Data               & Network data activity in kb/sec (total received, total transmitted, cellular received, cellular transmitted). \\
    & Light              & Mean light level in lux. \\
    & Noise              & Mean noise levels in dB. \\
    & Semantic Location  & Location visited by the user, classified as \emph{Home}, \emph{Work}, \emph{Out} and \emph{Unknown}.\\

    \hline
    \multirow{9}{*}{\rotatebox[origin=c]{90}{Event-driven}}
    & App                & Name and category of the app that was opened by the user. \\
    & Audio              & Change in the audio playback state (Music, No Music) or in the audio output of the device (Speaker, Headphones). \\
    & Charging State     & Charging state of the device (Charging, Not Charging). \\
    & Notification       & Post or removal of notification. \\
    & Notif. Center      & Event of accessing the device's notification center. \\
    & Ringer Mode            & Change of the ringer mode (Normal, Silent, Vibrate). \\
    & Screen             & Change of the device's screen state (On, Off, Unlocked). \\
    & Screen Orientation & Change of screen orientation (Portrait, Landscape). \\
    \hline
\end{tabular}
\caption{Periodical and event-driven sensor data collected by a smartphone device. For some sensors, mean and maximum information was aggregated by the app internally for power efficiency and respecting the user's privacy.}
\label{tab:sensors}
\end{table*}

% Weights
The input data for the classical approach consists of one instance per notification (notification entry or row). Each instance contains: the ground truth whether the notification was attended within the next 10~minutes after posting (binary), the aforementioned set of 145~features characterizing phone usage prior to the moment in which the notification was posted, and the relative weight of the instance. Weights are traditionally used by machine learning models to fight class imbalance~\cite{friedman2001elements}: instances with significantly fewer samples typically get higher weights to force the model into considering them equally. However, in our case, the weights are not only used to balance the different labels (notification attended / unattended), but also to balance the contribution of each app category and user. This way, we want to prevent that, for instance, apps or users with way more notifications than others dominate the training process. We consider 4~different strategies for compensating the aforementioned imbalances~\cite{Manning08BOOK} with an appropriate weight $w_i$ per instance $i$: (1) inverse frequency weighting, $w_i=1/f_{a,u,c}$, where $f$ is the frequency count of attended or unattended notifications ($a$) per user ($u$) and app category ($c$); (2) inverse square root frequency weighting, $w_i=1/\sqrt{f_{a,u,c}}$; (3) inverse log-frequency weighting, $w_i=1/\log(f_{a,u,c})$; and (4) no weights, $w_i=1$.

\subsubsection{Model}
% -------------------------------
As a learning model we use XGBoost~\cite{Chen16KDD}. XGBoost is a state-of-the-art gradient boosting regression tree algorithm that has emerged as one of the most successful feature-based learning models in recent machine learning competitions\footnote{\url{http://blog.kaggle.com/2017/01/23/a-kaggle-master-explains-gradient-boosting/}}. It produces a prediction model in the form of an ensemble of weak prediction models (single trees), and trains such ensemble following a stage-wise procedure, always under the same (differentiable) loss function~\cite{friedman2001elements}. In our case, we empirically found XGBoost consistently outperforming other well-established classifiers, such as logistic regression, support vector machines, or random forests.

\subsubsection{Technical Details}
% -------------------------------
We use XGBoost\footnote{\url{https://github.com/dmlc/xgboost}} v0.60. In pre-analysis, we found that only two configuration parameters had a noticeable effect on its performance: the maximum depth of the trees and the subsample ratio of the training instance~\cite{Chen16KDD}. Thus, we used grid search to optimize these two parameters on the validation set (see Section~\ref{sec:validationsets}). We also empirically configured XGBoost to use 101 estimators. Other than that, we use the default configuration and parameters.

\subsection{Deep Network Approach with Recurrent Neural Networks} 
\label{sec:rnn}
% ==================================================================

\subsubsection{Pipeline} \label{sec:rnnPipeline}
% -------------------------------
A deep feed-forward neural network~\cite{Goodfellow16BOOK} is a series of fully connected layers of units (nodes), capable of mapping an input vector (raw data) into an output vector (\emph{e.g.}, inferred classes). 
An RNN~\cite{Goodfellow16BOOK} is a specific type of network that takes sequential data as an input and, if implemented in a stateful way~\cite{Graves13ARXIV}, features an internal memory or state that allows it to remember past information. Notice that this is a very desirable feature in our scenario, as we want to have a kind of `summary' of what has been going on with the sensors at the moment of performing predictions (recall that in the classical approach above we empirically hard-coded this summary using hand-crafted features and a series of temporal windows prior to the notification posting).

\begin{figure}[t]
    \centering
    \includegraphics[width=0.9\columnwidth]{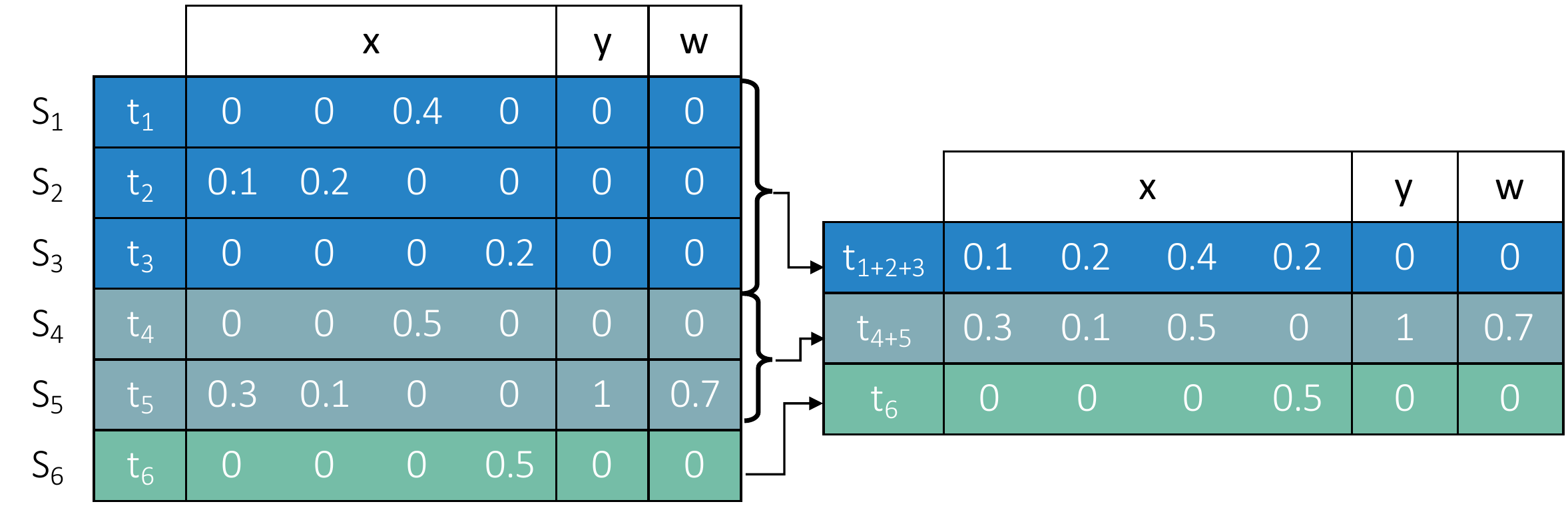}  
    \caption{Example of compressing sensor data.}
    \label{fig:compress}
\end{figure}

% Data organization
RNNs are typically designed for synchronous or re\-gu\-larly-sampled data. However, in our case, mobile sensors (specially the event-driven ones) yield sporadic and irregular events. Thus, we need to address the challenge of providing the sensor data in a form that can be effectively processed by an RNN. We organize the sensor event data in the form of a two-dimensional matrix: each row represents a sensor event $S_i$ and each column represents a sensor measurement $x$. 
For example, as illustrated in the left part of Figure~\ref{fig:compress}, in the sensor event $S_1$, a sensor writes its value into the third column. The next sensor reports two values, and writes them into columns 1 and 2. All other columns are set to 0. This way, we implicitly encode which sensor event has occurred. Ground truth labels are also represented as a column in the matrix ($y$). To prevent that the RNN learns from sensor events where no ground truth is available, the weight $w_i$ is set to $0$ when a ground truth label is not available. 

% Time compression
Because of the relatively frequent sampling used in the mobile sensor data, input sequences can be massive. To better deal with this large amount of data, we perform an opportunistic, lossless compression. In particular, consecutive input samples are combined when there is no clashing of information between them. An exception to the rule is the timestamp $t_i$, which we substitute by the time difference between both combined samples ($t_{i+1}-t_i$, called \emph{time delta} from now on). This way, we establish that data from a subsequent sample $S_{i+1}$ can be merged into an existing sample $S_i$ only if all of the following rules are valid for all given input sensors $j$:
\begin{itemize} %\compresslist
    \item $S_{i}[j] = 0$ or $S_{i}[j] = S_{i+1}[j]$; in other words, we can only merge the next sample into the current one if the current value is zero (no existing data) or is equal to the value of the following sample.
    \item $S_{i}[j]$ does not contain ground truth (\emph{i.e.} $w_{i} \neq 0$).%sample weight is not zero).
    \item The time delta between the merged samples is not larger than a threshold $T$ (we do not set $T$ in our experiments as our periodical sensors are configured to a fixed sampling rate of 10~minutes; therefore, we have at least one sample every $T=10$~minutes). 
\end{itemize}

%If any of the values per sample violates these rules, the sample is not merged. 
An example of time-based data compression is shown in Figure~\ref{fig:compress}. Samples 1--3 can be merged into a single sample. Sample 4 cannot be merged with the previous as one of the sensor value clashes with the merged sample. Sample 5 can be merged with 4. Sample 6 cannot be merged with 4 and 5, as 5 contains ground truth and we want to make a prediction at this point, without adding future features.

While this compression process results in a much denser input, there are some drawbacks. Firstly, a prediction is slightly delayed until a compressed sample has been generated (if required, smaller $T$ values can be used to shorten this delay). Secondly, the time information about the inter-arrival time of the compressed events is distorted. Finally, sensors that trigger multiple times with the same value can be compressed into a single event. However, performing a time-based compression presents a number of advantages that we believe outweigh the previous drawbacks:
\begin{enumerate} \compresslist
    \item Models train faster --- With smaller sequences we have less samples to feed into the neural network. If those samples keep the same information (as it is the case), the process results in faster training times with no performance drop.
    \item More effective past information --- By compressing longer time spans into smaller sequences we can feed more information into the RNN. This is important since the attention to past time spans of current RNN architectures is limited, a phenomenon known as the vanishing gradients problem~\cite{Pascanu12ICML}. 
    \item Compact sequences --- The sequence size is so small that we can even think of not deploying any further processing on the phone (including the deep network) and send that information to a server performing the remaining operations.
\end{enumerate}

\begin{figure}[t]
    \centering
    \includegraphics[width=1\linewidth]{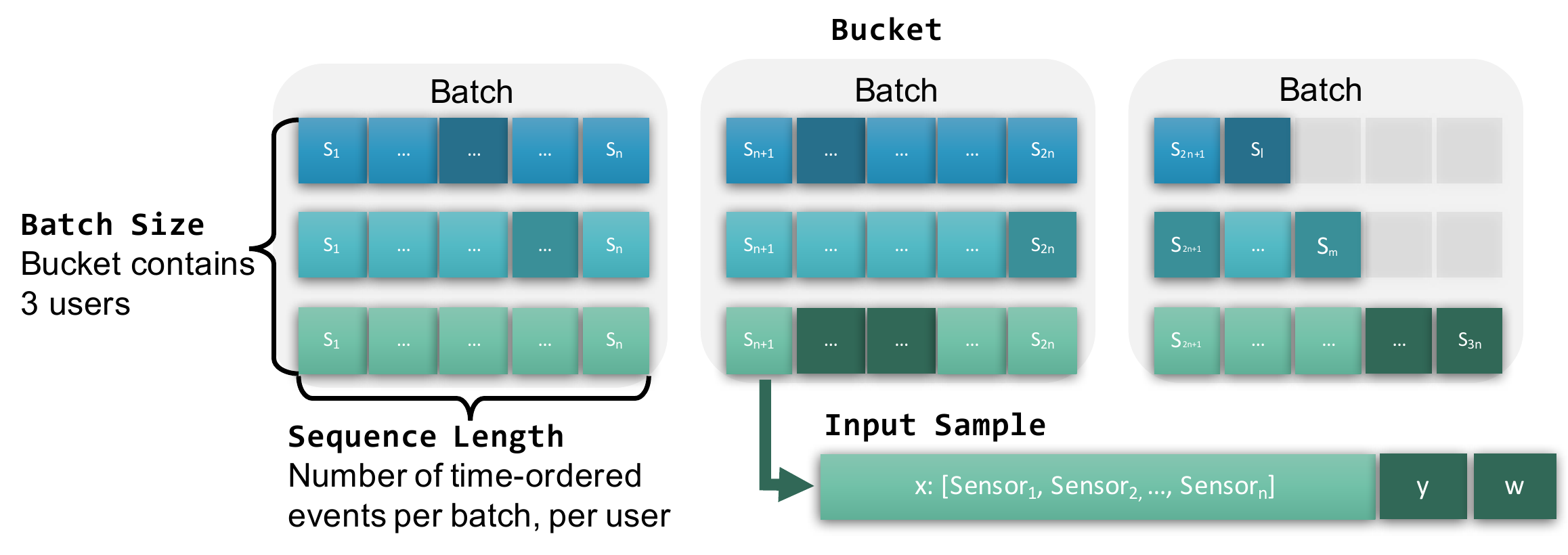}  
    \caption{Preparing data for training. Users are first split into buckets and then split into batches where multiple users are interleaved. Within each batch, a sequence of user data is provided. Some (but not all) of the samples contain the ground truth that will be used to train the model.}
\label{fig:batch}
\end{figure}

% Batching
If we take raw sensor events as an input, we find that some of the samples only contain sparse sensor readings and that, even after compression, not all of them contain labeled data or ground truth. Nevertheless, even if there are unlabeled sensor readings, all samples should go through the RNN, as these will keep updating the internal RNN state. We will train the network in this way, forcing it to perform a prediction at every input, even in the absence of ground truth. Therefore, we need a way to indicate that a given sample should be used to update the internal states (in other words, affect the past memory), but it should not be considered by the loss function in training time. To do so, we mark samples without a ground truth label with zeroed weights, whereas for instances that contain ground truth, weights are calculated following the same previous strategies introduced in Section~\ref{sec:xgboost}. In the example of Figure~\ref{fig:batch}, we see that the whole sensor input is passed through the network, but the network only learns from the highlighted samples. 

There are a number of data structuring operations that we perform before we can input the sensor data to the RNN. All these structuring operations are generic to RNN models, and totally automated without the need of human intervention. Such structuring concerns several dimensions (Figure~\ref{fig:batch}):
%\begin{itemize} \compresslist
    %\item
    
    \textbf{Input sample} --- Each sample $i$ contains the input data of a single instance for a single user: a tuple $S_i = (x_i,y_i,w_i)$, where $x_i$ is a vector of sensor events, $y_i$ contains the ground truth label, and $w_i$ contains the weight of this sample, used in the error or loss function. Notice that not all samples contain a ground truth label (see also Figure~\ref{fig:compress}).
    
    %\item
    \textbf{Sequences} --- To train RNNs, we need to provide, for each user, a time-ordered sequence of input samples. In our case, a sequence is a number of subsequent sensor events. These samples are used to build an internal state that determines how past events affect future time slots. They are also used to back-propagate the error when training the RNN~\cite{Goodfellow16BOOK}. The number of steps to perform this back-propagation in time, the \emph{sequence length}, is a parameter of the model.
    %\item
    
    \textbf{Batches} --- Modern techniques allow us to train a network in batches by interleaving multiple sequences, that is, data from multiple users, together. Among others, batching allows to further exploit the power of matrix multiplication on the graphical processing unit (GPU), and to avoid loading all data into memory at once. The batch size has some implications for the robustness of the error that is propagated in the learning phase~\cite{Keskar17ICLR}. Figure~\ref{fig:batch} shows an example of 3~batches that encode 3~sequences of 5~samples each (15~samples per batch in total). 
    
    %\item 
    \textbf{User buckets} --- By using stateful RNNs, the internal RNN state is kept between two subsequent batches, potentially allowing it to learn sequences that are larger than the sequence length. To do so, we need to make sure that two subsequent batches interleave the same users with the same order. Therefore, we assign them into buckets: each bucket contains all the batches that are required to encode the data of a group of users. If the users within a bucket have a different number of sequences, we zero-pad their data and sort them so that the minimum zero padding is needed. Figure~\ref{fig:batch} shows an example of a single bucket that encodes the data of 3 users. 
    
    %\item 
    \textbf{Prediction} --- Notice that the suggested arrangement into buckets and batches is only required in the training phase. For predictions we can even provide a single sample of a single user and the network will make a prediction based on the previous samples of that user, if available. In the case that no previous samples are available a `zero state' vector is passed to the RNN.
%\end{itemize}

\subsubsection{Model}  \label{sec:features}
% -------------------------------
We base our deep network model in a stacked architecture of long short-term memory (LSTM) cells~\cite{LSTM}. LSTM cells are the common recurrent architecture of choice in sequential tasks across a variety of domains~\cite{Goodfellow16BOOK}. At the very first stage, the input data is fed to a fully-connected layer of neurons with parametric rectified linear units (PReLUs)~\cite{He15ICCV} as nonlinear activations. The idea is to perform an initial embedding of the raw (compressed, but still sparse) data into a dense space, similarly to what is customarily done when processing text sequences~\cite{Graves13ARXIV}. Such dense representation goes through two stacked LSTM cells with sequential output, and the output of the second one is finally processed by another fully-connected layer with a sigmoidal activation that performs binary classification.

\subsubsection{Technical Details}
% -------------------------------
To implement our model we used Keras v2.0.3~\cite{keras} with Theano v0.9~\cite{theano} and, unless stated otherwise, we select the default parameters. We empirically set the number of neurons of the input layer to 50 and the number of units of each of the two LSTMs to 500. We train our model using standard cross-entropy loss~\cite{Goodfellow16BOOK} and the Adam optimizer~\cite{Kingma15ICLR}. We stop the training process when we see no improvement for 5 epochs and recover the best model seen so far in validation. We use a sequence length of 50 and a batch size that is automatically determined following a simple rule-based procedure: we select the smallest number between 15 and 45 that yields the minimum number of left-out users (that is, that yields the minimum modulo when dividing the number of users by the batch size).

The training of neural networks tends to converge faster if the input is normalized. Thus, before feeding the input into the network, we normalize it by re-scaling all the elements to lie between 0 and 1. To alleviate the issue of using 0 for both a missing value and a true 0 measurement, we re-scale data to range between 0.05 and 1. In sensor data, we typically find highly-skewed, long-tail distributions. We empirically tested different thresholds above which the values are capped, and ended up using the $95^\text{th}$ percentile of the input data. By capping the time delta value at 60~minutes, we also avoid outliers in situations like the device is switched off for some time or the battery runs out.

\section{Evaluation}
% =============================================================

\subsection{Metrics}
% -------------------------------------------------------------
As mentioned in Section~\ref{sec:xgboost}, our data set is unbalanced in three dimensions: users, notification categories, and notification attendance. If we want to compensate for the three imbalances at the same time, we need to go beyond plain evaluation metrics. One option would be to consider a weighting scheme for every ground truth item. However, this option would be subject to a correct choice of the weights, which is a decision we cannot make beforehand and, actually, it is a parameter whose influence we want to investigate. Moreover, we do not want such parameter to have an effect in the evaluation metric; we want such metric to be independent of weights.

Our strategy consists in separating the ground truth and predictions per user and per notification category. Then, for each combination of user and category, we compute the receiver operating characteristic (ROC) curve and, to have a single summary number, the area under the curve (AUC)~\cite{Manning08BOOK}. Once we have all the ROC curves and AUC values, we can then average them across categories, to have an average estimate for every user, and across users, to have a global estimate of the performance of the model.

Both the ROC curve and the AUC score compensate for the different number of attended versus unattended notifications~\cite{Manning08BOOK}. The ROC curve takes as an input a continuous prediction, such as the probability of a notification being attended, and a binary ground truth decision, such as the ground truth label. It then plots the true positive rate against the false positive rate at various threshold settings\footnote{The true-positive rate is also known as sensitivity, recall or probability of detection. The false-positive rate is also known as the fall-out or probability of false alarm, and can be calculated as 1 minus the specificity~\cite{Manning08BOOK}.}. The AUC is the total area under the ROC curve and ranges from 0 to 1, the latter representing perfect prediction.

\subsection{Validation Schema} \label{sec:validationsets}
\begin{figure}[t]
    \centering
    \includegraphics[width=0.9\linewidth]{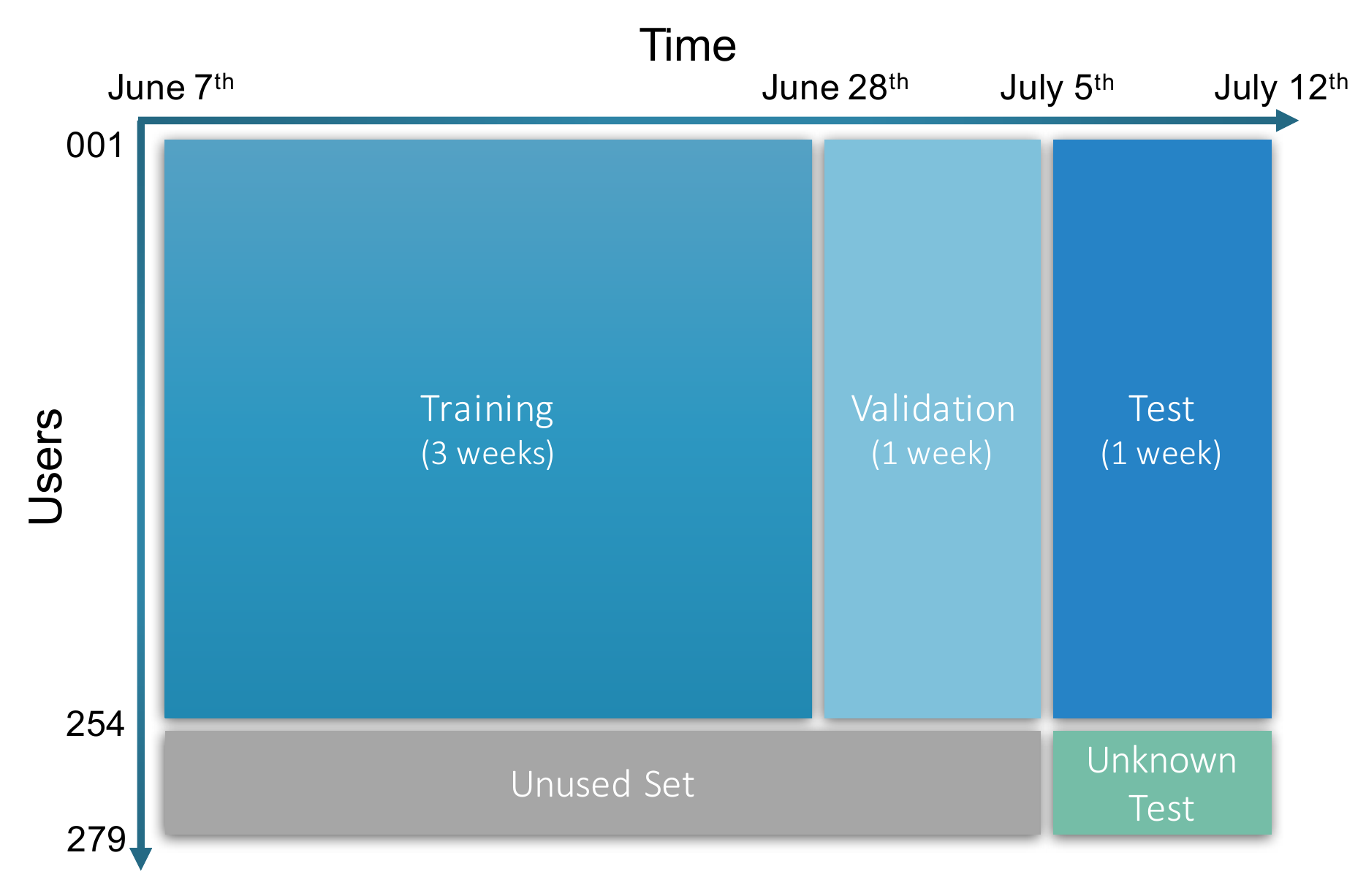}
    \caption{Diagram explaining the validation schema used in our analysis, featuring both time- and user-wise splits.}
    \label{fig:splits}
\end{figure}

% -------------------------------------------------------------
To train our models and perform a proper out-of-sample analysis, we split our 5-week sequential data set into training (first three weeks), validation (fourth week), and test (fifth week) sets (Figure~\ref{fig:splits}, horizontal axis). For the classical approach, the validation set is used to tune the aforementioned XGBoost parameters. For the deep network approach, the validation set is used to tune where to stop training, and as a reference measure to guide our empirical decisions regarding the number of neurons or the structure of layers. The unseen test set is used to report our evaluation metrics.

In order to have a further estimate of the performance of the models with unseen/new users, we generate an addition split from the test set, which we call \emph{unknown test}. Importantly, this split is not performed time-wise as the others, but user-wise and prior to the train/validation/test split (Figure~\ref{fig:splits}). We held out 25 users beforehand, chosen at random, and ensure that they do not enter the other splits. Then, to obtain a reliable estimate of the performance on all splits, as well as some indications of the variability of such performance, we perform 10 shuffle-and-split trials following the previous strategy. For every split, we report the average AUC and its standard deviation across trials.

\subsection{Baseline}
% -------------------------------------------------------------
As a baseline we use a probability-based random classifier. From the training set, it calculates the probability that a posted notification will be clicked within the next 10~minutes (for each user and each category individually). Then, in the prediction phase, it uses this probability as threshold in a random prediction. For example, if a user responded to 80\% of the WhatsApp messages within 10~minutes, the prediction will yield, on average, 80\% positive predictions. While a majority-based classifier is an equally common baseline, we chose a probability-based baseline as we observed a 5.09\% improvement in the F-score. Further, a majority-based classifier would not have allowed us to compute a proper AUC metric that considers multiple operating points as our classifiers do.
% Baseline: F = 0.530, Majority: F = 0.503

\subsection{Sensor Importance Calculation}
% -------------------------------------------------------------
Apart from assessing the general performance of the two approaches, it is important to evaluate how much each sensor contributes to make predictions about the attentiveness of a user towards notifications. While XGBoost provides a direct way to calculate feature importance, this is not possible in neural networks. Besides, feature importances for XGBoost should be aggregated per sensor, with no well-established, intuitive way of performing such aggregation. Therefore, in order to follow a consistent methodology between these two models, we use a greedy approach: we train multiple models, each time excluding the data of a single sensor (that is, each trained model contains 14 out of the 15~sensors described in Table~\ref{tab:sensors}). Finally, as before, we average the results over 10 runs with different seeds and compare with the model that contains data from all 15 sensors.

\section{Results} \label{sec:feature_analysis}
% ==================================================================

\subsection{General Performance}
% -------------------------------------------------
We first compare the general performance of the two approaches and the random baseline. As shown in Table~\ref{tab:general_aucs}, with an AUC of approximately 0.7, both approaches perform about 40\% better than the baseline. The classical and the RNN approaches perform similarly in all splits. The differences are not statistically significant (Wilcoxon signed-rank, $p>0.05$). The small difference between validation and test sets indicates that no overfit to train/validation data is present. The small difference between test and unknown test data sets suggests that generic, user-agnostic models, such as the ones we have considered here, extrapolate to unseen users for the task at hand (see below).

\begin{table}[t]
    \setlength{\tabcolsep}{5pt}
    \centering
    \begin{tabular}{l|ccc}
    \hline\hline
    \bf Approach & \multicolumn{3}{c}{\bf AUC} \\ 
                 & \bf Validation & \bf Test & \bf Test Unk. \\ 
    \hline
    Random      & 0.501 & 0.500 & 0.498 \\ 
    Classical   & 0.701 & 0.702 & 0.695 \\ 
    RNN         & 0.714 & 0.705 & 0.702 \\ 
    \hline\hline
    \end{tabular}
    \caption{10-run average of AUC scores for validation, test, and test unknown splits. Both classical and RNN approaches use inverse log-frequency weighting.}
    \label{tab:general_aucs}
\end{table}

\begin{figure}[t]
    \centering
    \includegraphics[width=1\linewidth]{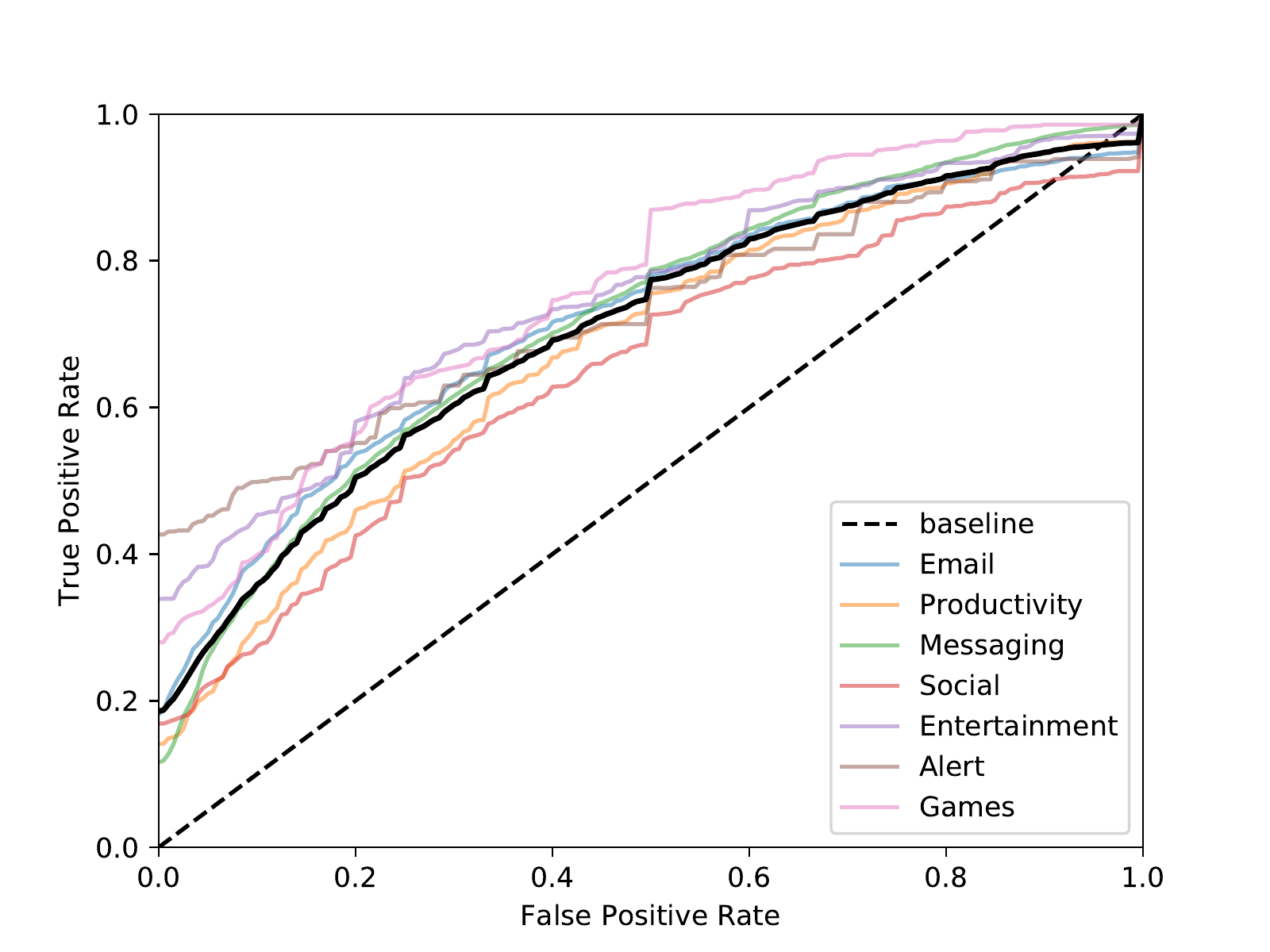}
    \caption{User-averaged ROC curves for a single run. The colored lines correspond to the considered app categories and the thick black line corresponds to the overall average across categories.}
    \label{fig:category_rocs}
\end{figure}

Figure~\ref{fig:category_rocs} provides a more fine-grained analysis of the results. It depicts the user-averaged ROC curves for every app category. The ROC curve shows the relation of sensitivity and specificity. Sensitivity in this context means: fraction of opportune moments that the algorithm identifies for the user to attend a notification. Specificity in this context means: fraction of correct predictions for the case that the algorithm predicts that it is not an opportune moment. Depending on the use case, applications can emphasize on one of these measures over the other. If it is more important not to hold notifications back by accident, such as in cases of messaging notifications, we can emphasize on sensitivity. If it is more important to avoid bothering users by accident, such as in the case of game notifications, we can emphasize on specificity.

We observe that, qualitatively, all curves present a similar shape. Attendance to social app notifications is found to be the most difficult prediction scenario, whereas alert and entertainment are the ones which are more easy to predict. Interestingly, with the latter two, we achieve true positive rates that are between 40 and 50\% with a very low false positive rate (below 10\%). In terms of sensitivity and specificity, and considering overall performance, we could build a classifier with 50\% sensitivity at 80\% specificity (or, if we want to be stricter, 35\% sensitivity at 90\% specificity).

\begin{figure}[t]
    \centering
    \includegraphics[width=1\linewidth]{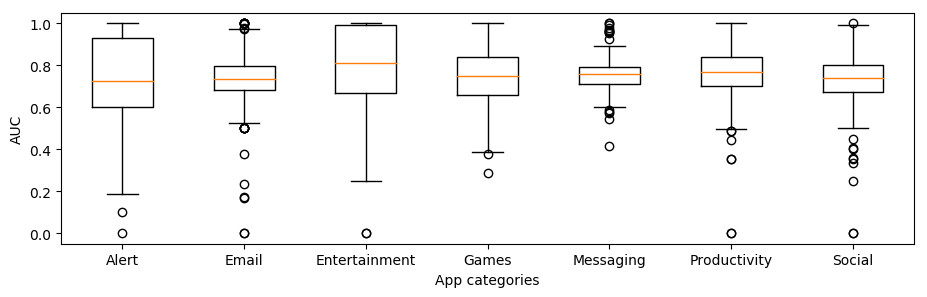} \\
    \includegraphics[width=1\linewidth]{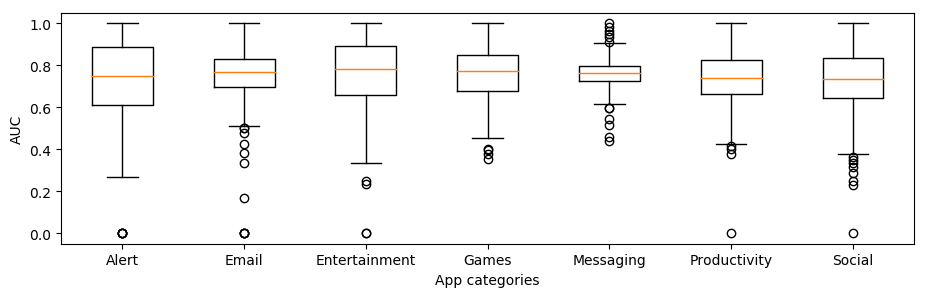}
    \caption{AUC box plots per category: classical approach (top) and RNN approach (bottom). The variability in the boxes corresponds to the considered users.}
    \label{fig:boxplots}
\end{figure}

Figure~\ref{fig:boxplots} visualizes the mean AUC by users, split by category, as boxplots.
We see that outliers are present, with users being easy to predict in some categories and then being complete outliers in others. It is interesting to note that, in both classical and RNN approaches, messaging presents the lowest spread among app categories. This indicates that all users are approximately equally predictable regarding this app category. Categories with the most spread are alert and entertainment. One explanation might be the low number of notifications in these categories.

\subsection{Sensor Importance}
% -------------------------------------------------

\begin{figure}[t]
    \centering
    \includegraphics[width=1\linewidth]{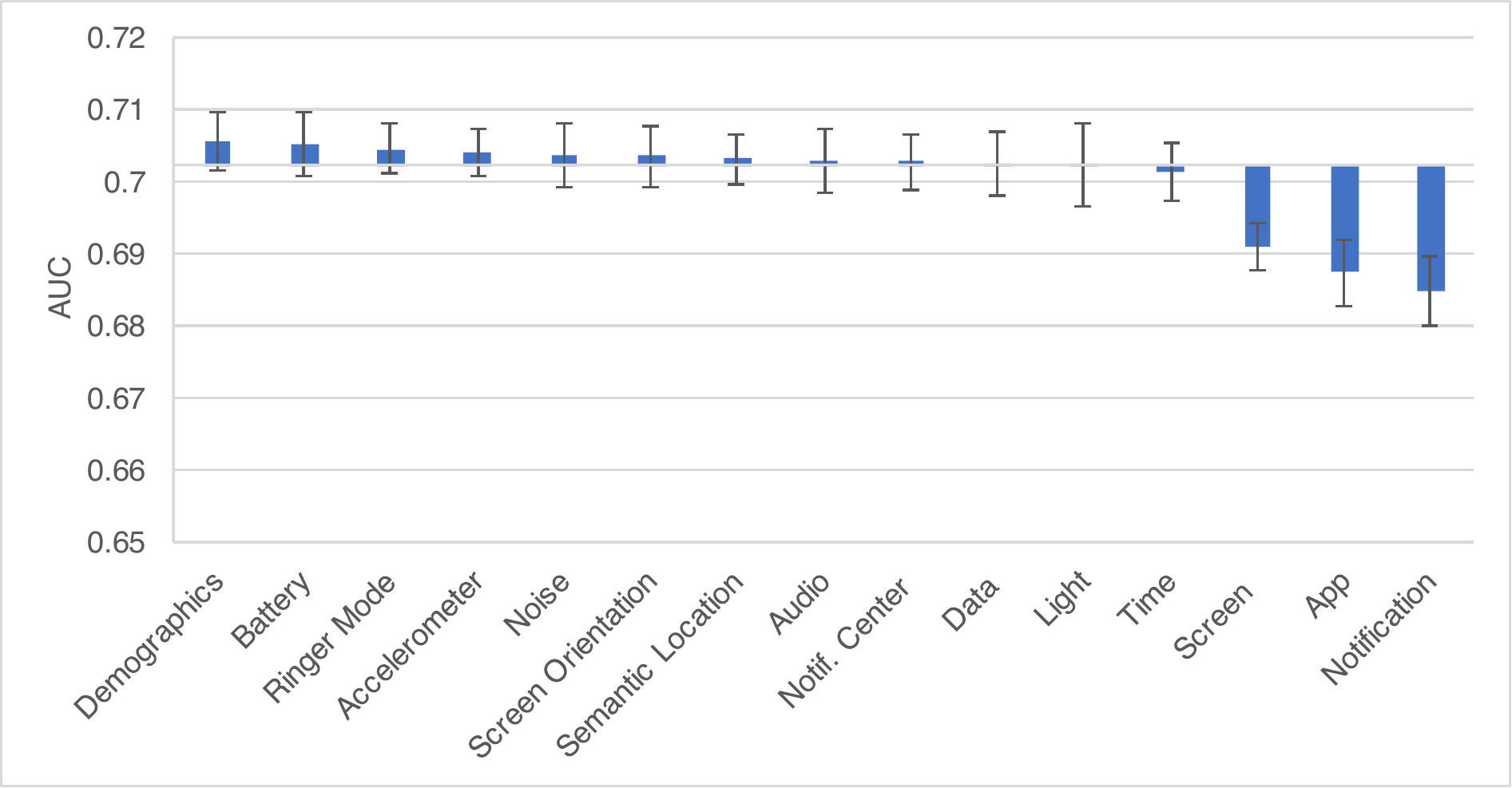}\\
    \includegraphics[width=1\linewidth]{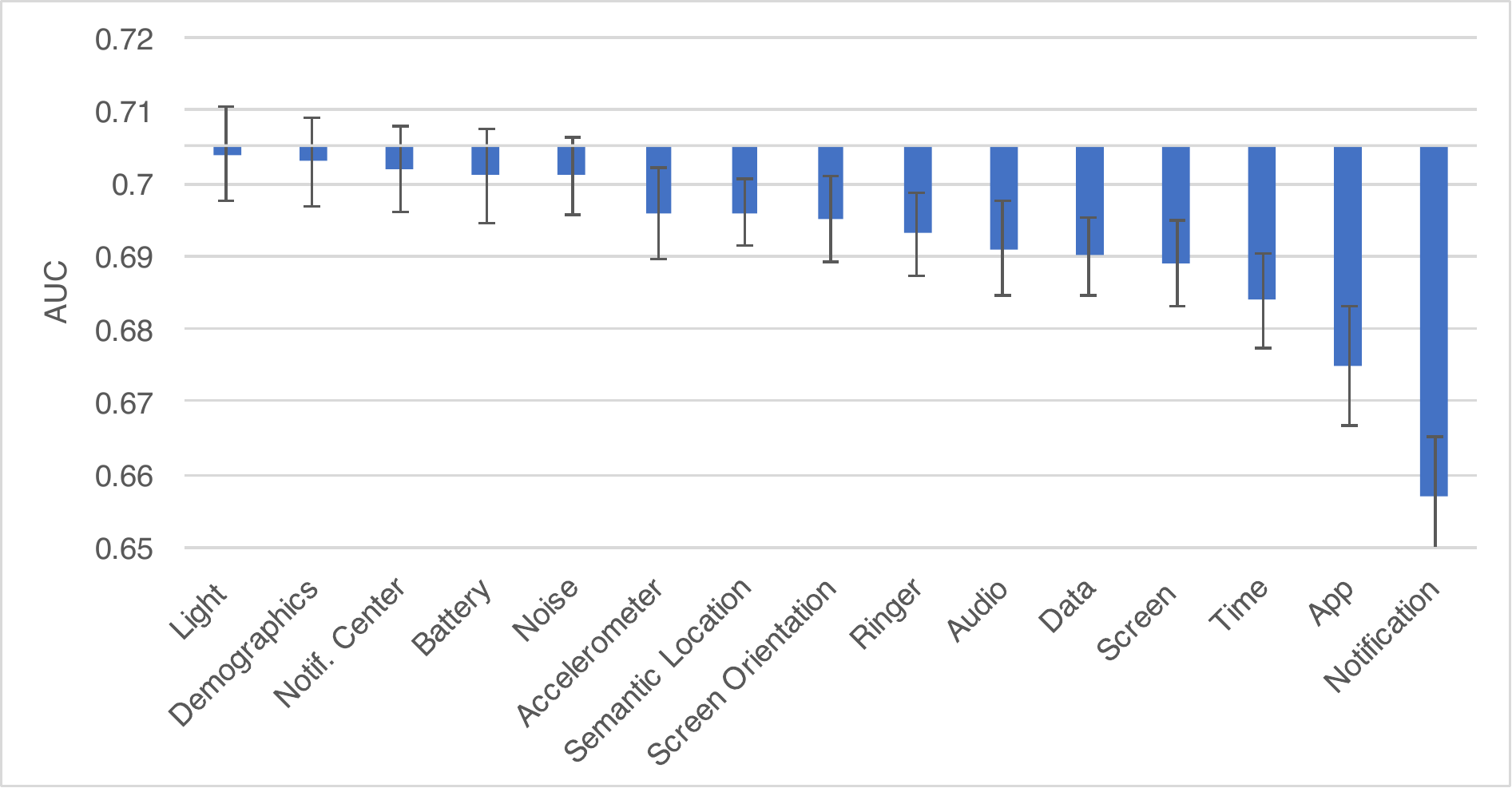}
    \caption{Feature importance quantified by mean loss of AUC when removing features of individual sensor: classical (top) and RNN (bottom) approaches.}
    \label{fig:featimp}
\end{figure}

As we observe in Figure~\ref{fig:featimp} (top), 
when we exclude individual sensors,
the performance of the classical approach drops notably for only three sensors:
i) the notification sensor ii) the application launched sensor and iii) the screen status sensor.
Surprisingly, we observe that individually removing any other sensor does not substantially affect the performance. This can be attributed to the fact that these sensors do not carry any information concerning proneness to handling notifications, or to the fact that multiple sensors may be correlated (that is, there is multicollinearity~\cite{friedman2001elements}).

The deep network model (Figure~\ref{fig:featimp}, bottom) shows similar behavior: the 3 sensors that are most important for the classical approach are also the important ones here. However, contrary to the classical approach, the performance of the RNN model is notably affected by the exclusion of additional sensors, such as the data usage, time of the day, ringer status, screen orientation, and semantic location. This indicates that the RNN is able to exploit more and make use of the additional information provided by the other sensors, being less dependent on a very select set of sensors. We hypothesize that the ability to exploit this information is why the RNN slightly outperforms the classical approach.

\begin{figure*}[t]
    \centering
    
    \includegraphics[height=3cm]{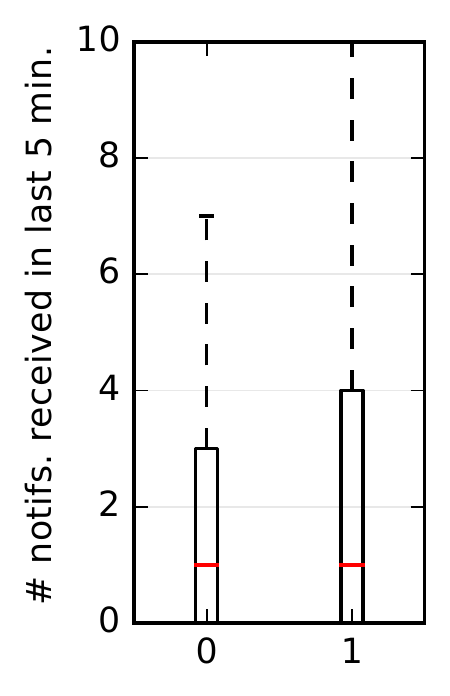}
    \includegraphics[height=3cm]{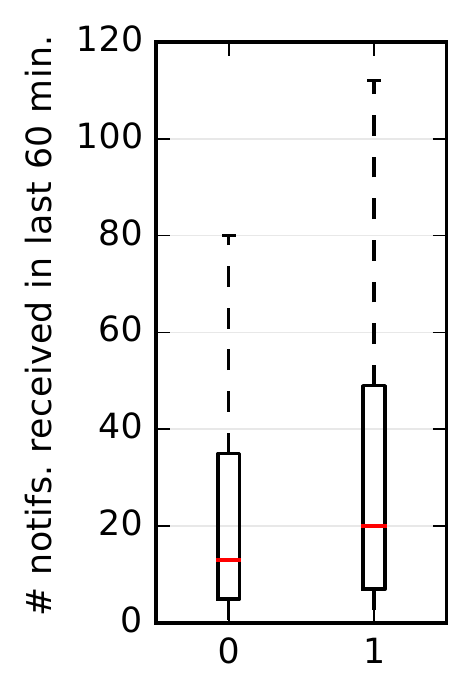}
    \includegraphics[height=3cm]{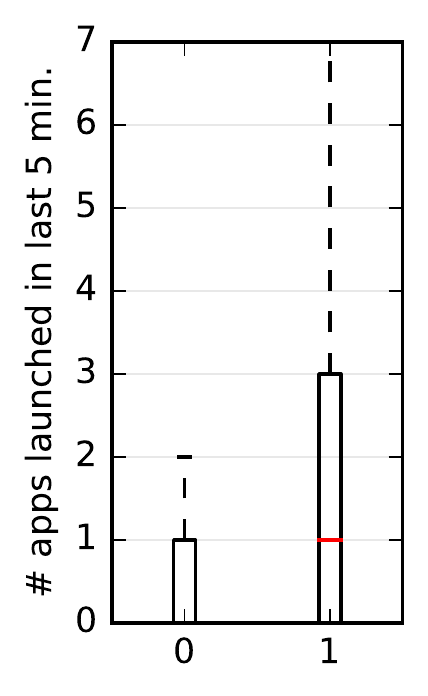}
    \includegraphics[height=3cm]{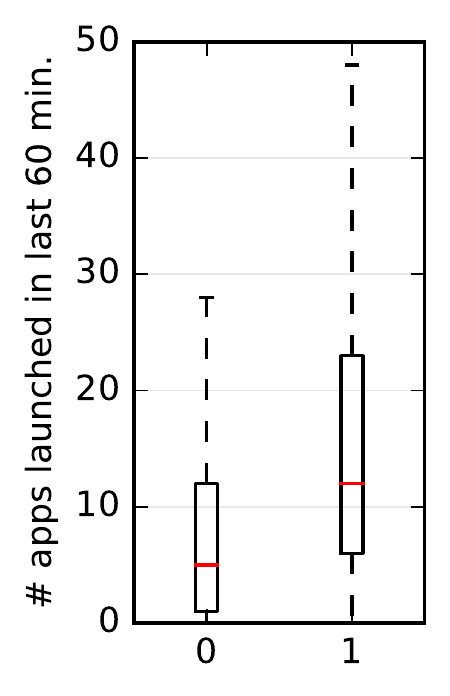}
    \includegraphics[height=3cm]{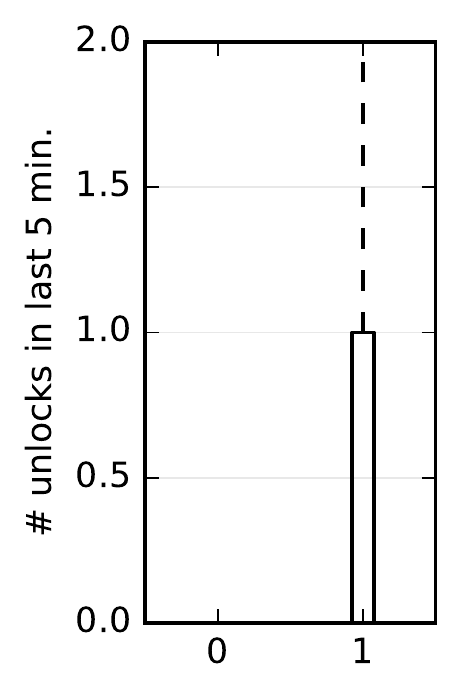}
    \includegraphics[height=3cm]{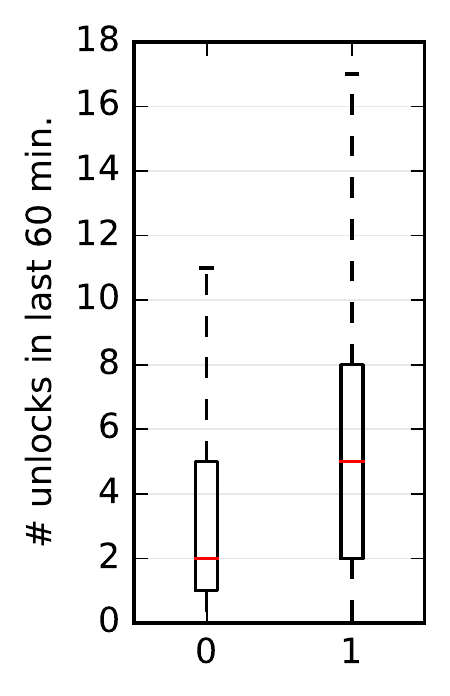}
    \includegraphics[height=3cm]{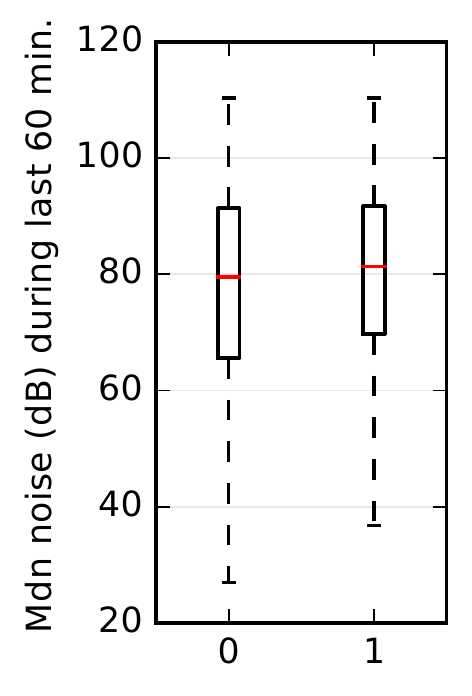}
    \includegraphics[height=3cm]{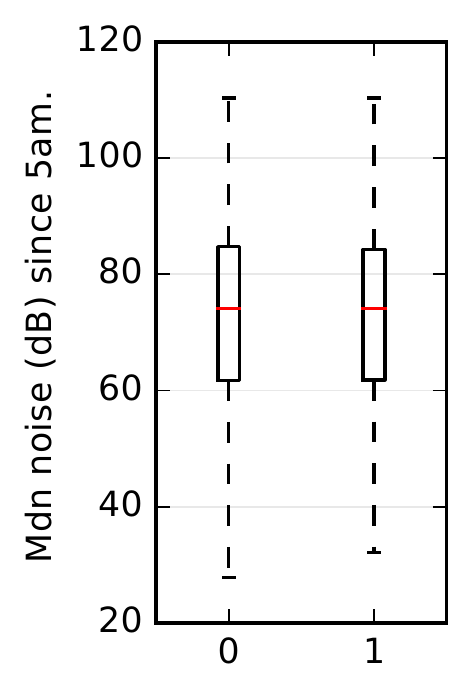}
    \caption{Comparison of feature values by ground truth for selected sensors. The figure shows the three most important sensors (screen sensor, app sensor, and notification sensor), as well as the average-performing noise sensor as reference. On the horizontal axes, 1 means notification clicked within 10 minutes, 0 otherwise.}
    \label{fig:featcomp}
\end{figure*}

According to above analysis, notifications, application launches, and screen activity appear to provide the most critical information about the interactivity and responsiveness of a user just prior to receiving a notification. Since one of our goals is to understand why these features better interpret attentiveness to notifications, we plot the distribution of these features for the two prediction classes in Figure~\ref{fig:featcomp} (we also plot the `microphone noise' feature to demonstrate these distributions when features have low prediction power).

As we observe in Figure~\ref{fig:featcomp}, users are much more likely to attend to a notification when they have received multiple notifications in the past 5 and 60~minutes. This can be attributed to the fact that multiple notifications might catch user attention, or the fact that the user is more interactive with his phone (for instance, chatting with someone). Finally, this feature might also capture other causes such as the time of the day  (for instance, less notifications are received at night).

The second important characteristic is the number and type of applications that the user has recently launched. As we observe in Figure~\ref{fig:featcomp}, the difference between the two classes is more pronounced: users that clicked on a notification have launched twice as many apps in the last few minutes. 
Similarly, users are much more likely to attend to a notification when they have unlocked their phone multiple times in the near past.  
Intuitively, these features capture the fact  that  users who have used their devices recently are more likely to keep using them in the near future. This can happen for instance because they are in general available (that is, non busy), or because they are constantly interacting with an application or someone. 
In contrast, the ambient microphone noise around the device just before receiving a notification does not correlate with the user clicking on it. We were hoping that this feature might capture a correlation between being at an open space (or on a conversation/meeting) and user attentiveness. However, this does not seem to be the case.

\subsection{Pipeline Decisions}
% -------------------------------------------------

In this section, we discuss the effect of the most important parameters in our pipelines. First of all, we assess the performance of both the classical and the RNN approaches with regard to the weighting strategy (Figure~\ref{tab:weight_effect}). We observe that, besides the simple inverse frequency weighting, all other strategies perform similarly. Inverse logarithmic weighting performed slightly better than the rest for both approaches.

\begin{table}[t]
    \setlength{\tabcolsep}{5pt}
    \centering
    \begin{tabular}{l|ccc}
    \hline\hline
    \bf Weighting & \multicolumn{3}{c}{\bf AUC} \\ 
         & \bf Validation & \bf Test & \bf Test Unk. \\ 
    \hline
    $1/f$         & 0.694 & 0.689 & 0.672 \\ 
    $1/\sqrt{f}$  & 0.702 & 0.702 & 0.696 \\ 
    $1/\log(f)$   & 0.701 & 0.702 & 0.695 \\ 
    $1$           & 0.706 & 0.698 & 0.709 \\ 
    \hline
    $1/f$         & 0.678 & 0.681 & 0.669 \\ 
    $1/\sqrt{f}$  & 0.704 & 0.700 & 0.700 \\ 
    $1/\log(f)$   & 0.714 & 0.705 & 0.702 \\ 
    $1$           & 0.714 & 0.702 & 0.702 \\ 
    \hline\hline
    \end{tabular}
    \caption{Effect of different weighting schemes: classical (top rows) and RNN (bottom rows).}
    \label{tab:weight_effect}
\end{table}

%Compression
The data compression for the RNN showed a number of notable benefits.
Firstly, as shown in Table~\ref{tab:compress_effect}, there is a big reduction (20x) on the data required to train and evaluate the model. This also results in significantly faster (28x) training times. The model training time improved significantly from 1.3~hours to 2.8~minutes per epoch on a device with a single GPU. 
Only thanks to the compression, the RNN achieved a prediction performance at par with the classical approach (AUC 0.705 compressed vs.~0.671 uncompressed on the test set). This is attributed to the fact that, by compressing longer time spans into smaller sequences, we can summarize more past information into the internal RNN memory state (see Section~\ref{sec:rnnPipeline}).

\begin{table}[t]
    \setlength{\tabcolsep}{5pt}
    \centering
    \begin{tabular}{l|ccc}
    \hline\hline
    \bf Compress & \multicolumn{3}{c}{\bf AUC} \\ 
                 & \bf Validation & \bf Test & \bf Test Unk. \\ 
    \hline
    No  & 0.678 & 0.671 & 0.660 \\ 
    Yes & 0.714 & 0.705 & 0.702 \\ 
    \hline\hline
    \end{tabular}
    \caption{Effect of applying compression to the RNN approach. Size reduction w.r.t.\ no compression was 20$\times$ and runtime speedup was 28$\times$.}
    \label{tab:compress_effect}
\end{table}

%Normalization and Capping
As a final note, we shall mention that data normalization and capping were important to make the RNN model converge faster. Without this, the distributions of the inputs was unbalanced. Thus, the learning rate was not able to make the required corrections at each iteration. This resulted into unstable loss results per epoch and, consequently, convergence issues while training the model.

\section{Discussion and Implications} \label{sec:discussion}
% ==================================================================

\subsection{Prediction Performance}
% -------------------------------------------------

% SUMMARY
We provide evidence that a model of past phone usage logs can predict whether a user will launch the app associated to a given notification within 10~minutes. Such model is trained using notifications from multiple app categories and requires no user input. Interestingly, its predictions perform equally well across all categories. This is somewhat surprising given that notifications of different categories are handled by humans in quite different ways. In particular, messages are handled a lot faster; presumably because, in contrast to other types of notifications, they are often associated with the pressure to respond timely~\cite{Pielot:2017}. 

Previous work has argued that content and sender-receiver relationship are the most important indicators for how a notification will be handled~\cite{Fischer2011,Mehrotra2015}.
When asking mobile phone users, they report that they click on notifications because the sender is important or because its content is important, urgent, or useful~\cite{Mehrotra2015,Mehrotra2016}. However, our findings show that a classifier that is agnostic to these factors performs surprisingly well. The most important metrics were related to how much the phone was in use, namely, notifications, app launches, and screen activity. This confirms previous findings that usage intensity predicts how fast mobile phone users will attend to a message~\cite{Pielot2014}, as well as openness to proactively recommended content~\cite{Pielot2015}. It further confirms the finding that people respond slowest to notifications when they self-report themselves as idle~\cite{Mehrotra2016}.

As to why the prediction works so well, we hypothesize that the intensity of phone use might correlate with \emph{ritualistic} phone use, that is, ``to browse, explore, or pass the time''~\cite{Hiniker2016}. Higher usage intensity was found to correlate with engagement~\cite{Mathur2016}, which we hypothesize represents a state of being more open to notifications. This is supported by the finding that if tasks are not challenging or important, mobile phone users self-report themselves to be more open to interruptions~\cite{Pejovic2015}.
This consideration leaves us with an interesting open question related to studying interruptibility through self-reports: to what extent is there a disconnect between what people think about interruptibility versus how they actually react to notifications?

\subsection{Beyond the Results: Combination and Personalization}
% -------------------------------------------------
The fact that both classical and RNN approaches achieve the same result, much better than the baseline, could lead one to think that they have reached the best possible performance score for this specific task and data set (pointing to what in machine learning folk-knowledge is known as a `glass ceiling'~\cite{pachet2004improving}). However, in our case, a close inspection to the scores of the two approaches on the same individuals reveals that they do not produce exactly the same errors (compare, for instance, the AUC score outliers in the box plots of Figure~\ref{fig:boxplots}). This indicates that there might be the possibility of combining the results of the two approaches to achieve a better final score. This line of work, including the research of whether it is better to perform an early or a late fusion (and how to perform it)~\cite{alkoot1999experimental}, is out of the scope of the present work, and remains an open question for future investigations.

Another aspect worth mentioning is the possible effect of personalization in our models. Here, we have not considered any personalization strategies, tackling the problem with generic, user-agnostic models. Nonetheless, we have studied the performance of the models on new/unseen users (the unknown test set), and we have seen that the scores obtained with these unknown users are comparable to those achieved with the known users used for training. This suggests that such generic, user-agnostic models may work well in practice with unseen/new users, effectively dealing with the so-called cold start problem~\cite{Manning08BOOK}. Adding a personalization aspect may improve performance with known users, but would not presumably change the scores with unseen/new ones, as we would not have the data to make such personalization reliable and effective.

\subsection{Pipelines' Comparison}
% -------------------------------------------------
Performing a qualitative comparison between the presented classical and the RNN approaches, and based on our experience, we note that each of them required a different pipeline. We now summarize their main conceptual differences: % (Table~\ref{tab:comparePipelines}):

% \begin{itemize} \compresslist
    %\item
    \textbf{Feature construction} --- While the classical approach required domain expertise to design meaningful features, the RNN did not require such construction. 
    %\item
    
    \textbf{Time aggregation} --- 
    Traditional machine learning approaches require aggregating temporal information over one or several time windows. In our implementation, we constructed features such as the ``number of unlocks in the last $T$ minutes'' and used multiple window sizes for each sensor to construct features capturing all relevant time spans. The selection of window sizes was based on domain knowledge%~\cite{Pielot2017}
    , but it could also be treated as a hyper-parameter to be optimized during training. Alternatively, features can be constructed based on minutes since an event has triggered. The RNN, on the other hand, did not require such aggregation as it can automatically identify potential temporal patterns.
    %\item
    
    \textbf{Training input representation} --- For the classical approach, each input sample had to contain the ground truth (the user handling a notification), enriched with the constructed features. In contrast, the RNN approach did not require such pre-processing. Each input sample was an independent sensor event that just updated an internal state. As far as the RNN is concerned, handling or ignoring a notification is just another sensor input. While the internal state is updated at every input, the network parameters are updated exclusively after handling ground truth events.
    
    %\item 
    \textbf{Prediction} --- Similarly, to make a prediction at any given time, the classical approach required to generate all the features based on the temporal windows as opposed to RNNs, where continual predictions were made whenever there was a new sensor reading.
    
    %\item 
    \textbf{Training speed} --- While there was a cost to construct the features for each notification, training with XGBoost was considerably faster (minutes vs.\ hours). Furthermore, since features were already provided, XGBoost converged to a solution within much less time. 
    
%\end{itemize}

\subsection{Server vs.~On-device Deployment}
% -------------------------------------------------

Due to the applied data compression, we employed a server-based implementation for both training and prediction. Apart from this being feasible with modern data connections, there are advantages too: the existence of centralized data allows us to train the model using ever-increasing multi-user input, the processing power allows us to  update the classifier in real-time making it easier to build and maintain complex personalized models. However, running the model on-device might be desired too as no network data would be consumed, there would be little delay between data collection and the decision and, most importantly, the users' privacy would be preserved.

\section{Conclusions} \label{sec:conclusions}
% ==================================================================

We demonstrate how to use an RNN to predict directly from raw mobile phone use events whether a posting a notification would lead to a launch of the associated app within the next 10 minutes. We provide evidence that this RNN performs at par with (if not better than) a competitive feature-based classifier.

% APPLICATION
To the best of our knowledge, this is the first work to make predictions on implicit behaviour towards actual notifications using RNNs. For training, the classifier does not require explicit user input, such as self-reported interruptibility. This enables intelligent ways to make notifications less disruptive and annoying. For example, when the user is not predicted to open the app soon, three major applications strategies could be applied: (i) suppress the notification, \emph{i.e.}, deliver it silently or, if not important, not deliver it at all;
(ii) defer until opportune moment, \emph{i.e.}, delay the notification until the user is predicted to be likely to open the app and post it then; and (iii) communicate unavailability, \emph{i.e.}, in the case of computer-mediated communication, inform the sender that the message is not likely to receive a response and therefore relieve the receiver from the pressure to respond timely.

% FUTURE WORK
Future work includes the exploration 
of more sophisticated techniques, such as transfer learning or unsupervised learning with the use of generative adversarial networks to further improve the prediction performance. 
An additional line of research could be the fusion of both classical and RNN approaches. In such research, feature level (early) or classifier level (late) fusion could be compared (\emph{e.g.}, input the features used for the classical approach to the RNN, or employ a meta-classifier to learn the best combination between outputs of the two approaches).

\section{Acknowledgments}

The authors wish to thank the participants of the study. They also wish to thank Alexandros Karatzoglou for the useful discussions.

\bibliographystyle{abbrv}
%\small
%\footnotesize
%\balancecolumns
\bibliography{deepnotifs,deepnotifs_joan}

\end{document}